\documentclass[]{spie}  
\usepackage[]{graphicx}

\title{The Commissioning Instrument for the Dark Energy Spectroscopic Instrument} 

\author{Ashley J. Ross\supit{a}, Paul Martini\supit{a,b}, Rebecca Coles\supit{b}, Mark Derwent\supit{b}, Klaus Honscheid\supit{a,c}, Thomas P. O'Brien\supit{b}, Dan Pappalardo\supit{b}, Suk Sien Tie\supit{b}, David Brooks\supit{d}, Michael Schubnell\supit{e}, Greg Tarle\supit{e}
\skiplinehalf
\supit{a}Center for Cosmology and AstroParticle Physics, The Ohio State University, Columbus, OH 43210, USA; \\
\supit{b}Department of Astronomy, The Ohio State University, Columbus, OH 43210, USA\\
\supit{c}Department of Physics, The Ohio State University, Columbus, OH 43210, USA\\
\supit{d}Department of Physics \& Astronomy, University College London, Gower Street, London, WC1E 6BT, UK\\
\supit{e}Physics Department, The University of Michigan, Ann Arbor, MI 48109, USA
}

\authorinfo{Further author information: (Send correspondence to A.J.R.)\\A.J.R.: E-mail: ashley.jacob.ross@gmail.com\\Presented at SPIE June 2018; Proc. SPIE 10702, Ground-based and Airborne Instrumentation for Astronomy VII, 1070280}

\begin{document} 
  \maketitle 

\begin{abstract}
We describe the design of the Commissioning Instrument for the Dark Energy Spectroscopic Instrument (DESI). DESI will obtain spectra over a 3 degree field of view using the 4-meter Mayall Telescope at Kitt Peak, AZ. In order to achieve the required image quality over this field of view, a new optical corrector is being installed at the Mayall Telescope. The Commissioning Instrument is designed to characterize the image quality of the new optical system. The Commissioning Instrument has five commercial cameras; one at the center of the focal surface and four near the periphery of the field and at the cardinal directions. There are also 22 illuminated fiducials, distributed throughout the focal surface, that will be used to test the system that will map between the DESI fiber positioners and celestial coordinates. We describe how the commissioning instrument will perform commissioning tasks for the DESI project and thereby eliminate risks.
\end{abstract}



\section{INTRODUCTION}
\label{sec:intro}  

The Dark Energy Spectroscopic Instrument (DESI\cite{DESI_INST,DESI_SPIE}) project represents a huge advancement in the ability to obtain optical spectra. DESI will be used to conduct a survey to obtain spectra of 35 million galaxies and quasars. DESI is classified as a Stage IV dark energy experiment due to the high precision of its forecasted constraint on the dark energy equation of state\cite{DESI_SCI}. 

DESI will use robotic fiber positioners to simultaneously align 5,000 optical fibers with pre-defined targets distributed over a three degree field of view. The DESI ``Focal Plane System'' (FPS) \cite{DESI_FPS} comprises the combination of components to be installed at the prime focus of the Mayall 4 meter Telescope at Kitt Peak, AZ. In order to achieve the required image quality over the full three degree field of view, a new optical corrector will be installed on the Mayall in Summer 2018\cite{DESI_CorrA,DESI_Corrlens}.

The Mayall telescope is currently being prepared for the installation of the new optical corrector. The installation is scheduled to be completed in the summer of 2018. The DESI FPS is scheduled to begin installation in March 2019, followed by a period of commissioning and science verification.

The DESI timeline thus includes a several months between installation of the new corrector and and the time when the FPS is ready. This schedule gap, along with the absence of on-axis imaging with the FPS, motivated the construction of the Commissioning Instrument. It is being designed and fabricated so that it will be installed on the Mayall Telescope in the fall of 2018, soon after the new optical corrector is fully installed. The CI will commission and characterize the new optical system and the DESI Instrument Control System\cite{DESI_ICS} (ICS) software. Key components include the Telescope Control System (TCS), the Active Optics System (AOS), and the guider. Further, the CI includes an on-axis imager and thus will uniquely provide the DESI project with simultaneous direct measurements of the image quality at both the center and edge of the field of view.

The benefit of testing DESI systems with on-sky observations was demonstrated previously with the ProtoDESI\cite{ProtoDESI} experiment, which successfully demonstrated integration between the hardware, the DESI ICS, and the Mayall TCS. ProtoDESI also demonstrated the successful acquisition of targets and that these targets were held in place using the guide camera hardware and software.
The program further demonstrated that astronomical targets could be acquired using robotically positioned fibers and that the DESI guiding requirements could be met. The DESI CI will compliment this effort by testing DESI control systems and hardware with an observing campaign that will complete DESI commissioning tasks related to image quality and performance over the full field of view.

In this manuscript, we describe the design of the DESI Commissioning Instrument in Section \ref{sec:design} and the observing plan in Section \ref{sec:obs}. We conclude in Section \ref{sec:conclude}.

\section{DESIGN} 
\label{sec:design}

\subsection{Cameras and Fiducials}

The CI requires radial and on-axis imaging. The radial imagers will determine the image quality at approximately the same focal surface positions as the DESI Guiding, Focus, and Alignment (GFA\cite{DESI_GFA2}) cameras. The focal surface is curved, with a slope of over 5 degrees at its edge; the curved focal surface is accounted for in the design. The on-axis imaging will characterize the image quality beyond what will be possible with only GFA cameras on the outer diameter of the DESI Focal Plane System and thus provide imaging tests that more completely sample the DESI optical corrector.

The DESI CI uses five cameras. In order to properly image enough stars at any celestial location, the field of view of each CI CCD is required to be at least 25 square arcminutes and the pixel size less than 0.4 arcseconds. Additionally, a goal for the project is to enable guiding updates at a 10 second cadence and this makes a fast readout rate a priority. These requirements, time constraints, and our modest budget drove the decision to choose the commercially available SBIG STXL-6303e camera. This camera's CCD grid and pixel sizes (3072 x 2048 and 9 microns) correspond to a 28 square arcminute field of view covered by 0.13 arcsecond pixels. The quoted full frame download time of 4 seconds (1.8 MHz readout rate) was superior to comparable options. Further, we could mount an r-band filter in each camera, in order to match the transmission expected for the DESI GFA cameras\cite{DESI_GFA2}.

After we nominally chose the SBIG STXL-6303e, one was purchased for testing purposes. Testing the camera in the lab, we found a gain of 1.6 $e^{-}$/ADU and a readnoise of 13.7 $e^{-}$/pixel. Both quantities were acceptable given our imaging requirements and consistent with previous studies\cite{SBIGpaper}, though the readnoise is greater than the 11 $e^{-}$/pixel quoted by the manufacturer\footnote{http://diffractionlimited.com/product/stxl-6303e/}. Further, we found the dark current varied from 0.05 ADU$s^{-1}$ with the cooling set to 0 C to 0.38 ADU$s^{-1}$ with the cooling set at 20 C; factoring in the gain, these values are consistent with the 0.5 $e^{-}/s$ quoted by the manufacturer. This suggests the dark current will be at an acceptable level for CI operation when it operates at the ambient temperature of the Mayall Telescope dome, and we do not expect to require precise photometry for low surface brightness objects. Based on these tests, we chose the SBIG STXL-6303e.

Twenty two illuminated fiducials are placed throughout the CI focal surface to map the optical distortions of the corrector. These illuminated fiducials are the same model used for the full DESI Focal Plane System\cite{DESI_FPS} and are produced by Yale University\cite{ProtoDESI}. Their distribution for the CI is shown on Fig. \ref{fig:CIFPA}. Each illuminated fiducial contains four pinholes, which are illuminated by an LED. These pinholes are imaged by the DESI Fiber View Camera (FVC). This hardware and the software was demonstrated to to work during the ProtoDESI campaign\cite{ProtoDESI}. The observed positions of the pinholes allow for the mapping of the distortion pattern over the full field of view, and thus between $X,Y$ on the focal surface and $\alpha,\delta$ positions of targets on the sky. Recently developed methods\cite{Kent17} demonstrate that spin-weighted Zernike polynomials can be used to efficiently model optical distortion patterns, and that at most 16 coefficients will be required for DESI. 

We chose the locations of the 22 fiducials to produce a good measurement of the corrector distortion. This measurement requires accurate knowledge of the physical $X,Y$ locations of the illuminated fiducials. In order to accurately place the 107 micron fibers on their targets with an RMS uncertainty of 0.05 arcseconds, the relative $X,Y$ positions of the illuminated fiducials must be known to within 10 microns. We will also compare the astrometric solution obtained from CCD images to the optical model of the corrector and to the measurements from the illuminated fiducials. In order to minimize the uncertainty in the transformation between the measurement of the corrector distortions from the fiducials and the astrometric data from the CCDs, we require that we will measure the relative $X,Y$ position of the CCD and the nearest illuminated fiducial to within 5 microns. Achieving this accuracy is challenging, and our metrology procedure is described in a companion SPIE proceeding.\cite{ColesMetrology}

In Section \ref{sec:obs}, we will describe how the combination of the CI cameras and illuminated fiducials will be used to commission the DESI optical system and, in particular, map between $X,Y$ and celestial coordinates to the accuracy required for the placement of DESI optical fibers on astronomical targets.

\subsection{Mechanical Design}

The CI will commission the Active Optics System that maintains the alignment of the Corrector and Focal Plane System (or Corrector and Commissioning Instrument) with the primary mirror. Flexure of the telescope structure is probably the primary source of misalignment, and we have therefore matched the CI mass to the 815 kg mass of the Focal Plane System to mimic the expected behavior of the latter. The overall diameter is 1420 mm and the height is 1432 mm. Another key component is that the cameras and illuminated fiducials are placed in stable mounts and properly aligned with the nominal aspheric focal surface with less than 50 micron uncertainty relative to the surface normal.

The CI has three large, precision-machined steel components that mate to the corrector. There is also an additional weight of about 211 g that mimics the mass of the Focal Plane System thermal enclosure, and is attached to the cage. Steel was chosen as it is stiff, heavy enough to match to the mass of the full DESI Focal Plane System, and cost effective. 

The first component is the Adapter Disk. This is a steel ring that will be mated to the barrel of the optical corrector via a precision piloting feature.
 The origin of the counter-bored ring is offset from the full ring by -170,42 microns in $X,Y$. This offset was carefully determined with CMM metrology of the corrector barrel\cite{DESI_parts}, and the design of the CI thus corrects for this offset. Tapped holes allow the other two pieces of the CI to be mated to the adapter disk.

\begin{figure}
   \begin{center}
   \begin{tabular}{c}
   \includegraphics[height=7cm]{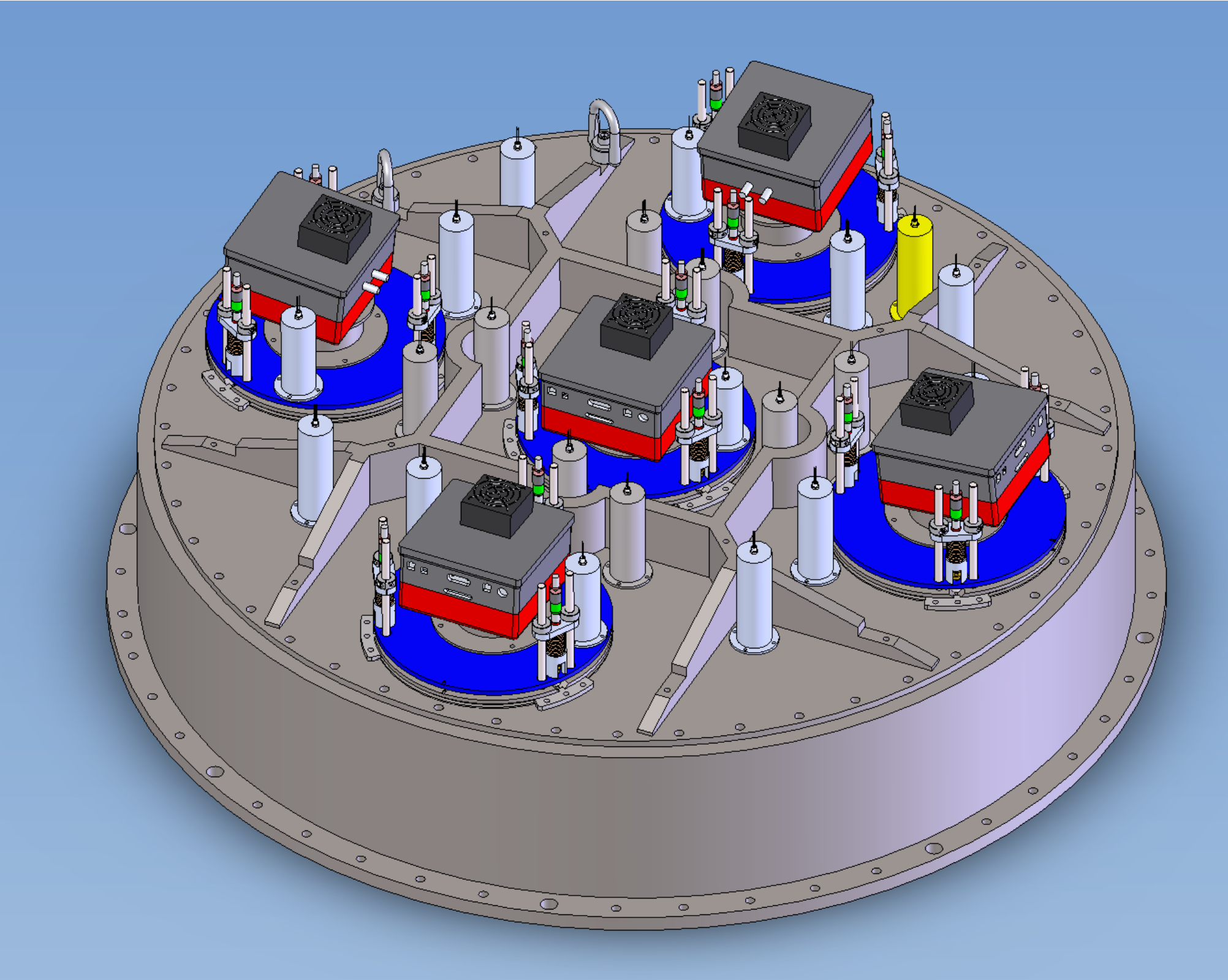}
   \end{tabular}
   \end{center}
   \caption
   { \label{fig:CIFPA} 
An image of the Commissioning Instrument Focal Plate Assembly. This is the backside of of the focal surface; light enters the cameras or exits the illuminated fiducials from below. It consists of a ribbed steel plate, five cameras that are mounted with their CCDs aligned to the focal surface, and 22 illuminated fiducials (one of which is shown in yellow).}
   \end{figure}

The second component is the CI Focal Plane Assembly (FPA). The CI FPA is made of two large pieces of machined steel. The first is a cylinder with a height of 23 cm that has an outer ring of tapped holes that mate with the Adapter Disk, and an inner ring of tapped holes. The height of this cylinder places the second piece of the CI FPA at the focal surface of the corrector. The cylinder also provides stiffness for the FPA. The rings are machined to be concentric to within 20 microns of each other and the adapter disk. The second piece is a ribbed plate with tapped holes that are bolted to the cylinder. Adjustable mounts (described below) for the cameras and illuminated fiducials are clamped into place on the plate. The ribs of the plate optimize its stiffness and mass. An image of the full CI FPA is shown in Fig. \ref{fig:CIFPA}.

\begin{figure}
   \begin{center}
   \begin{tabular}{c}
   \includegraphics[height=7cm]{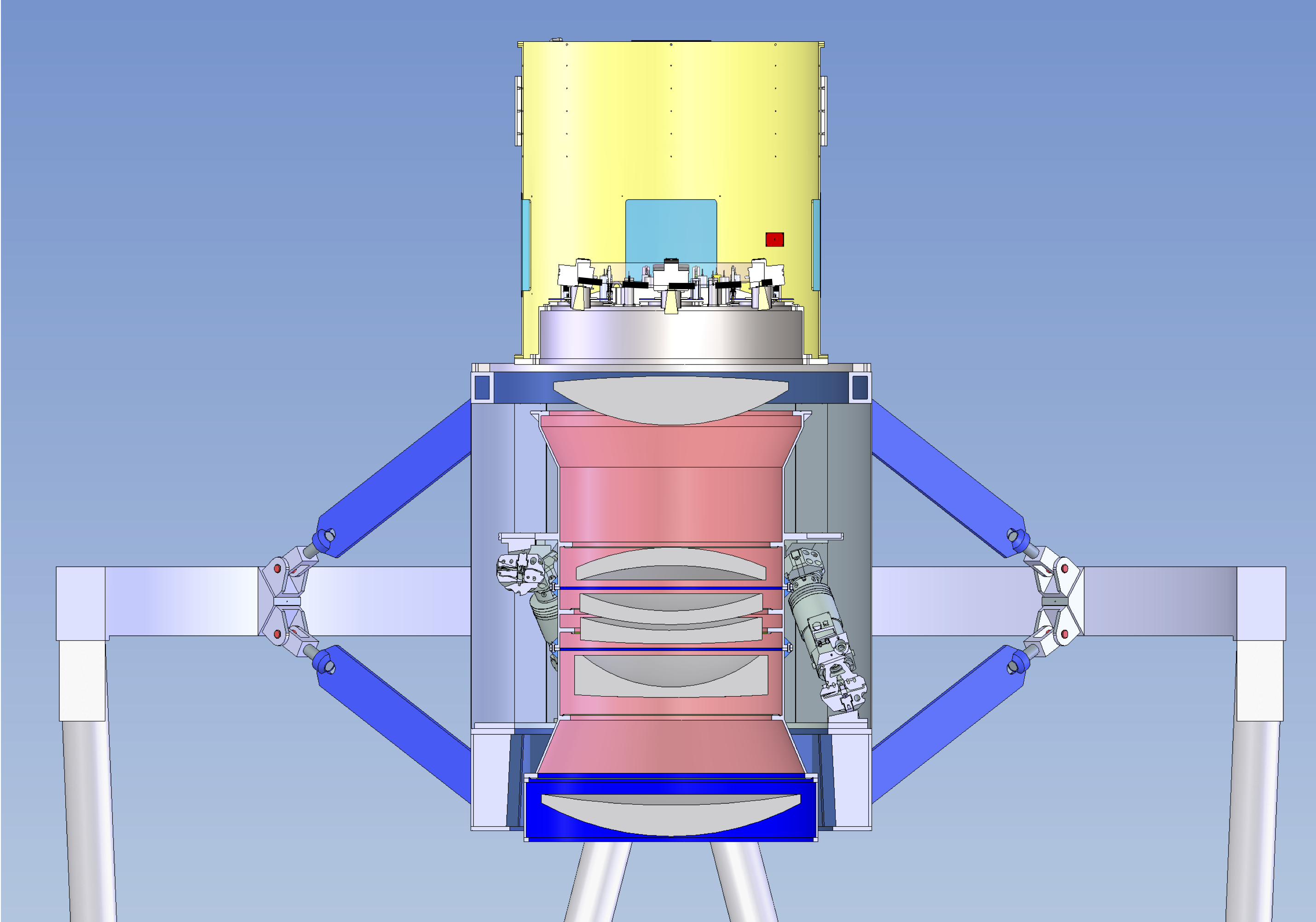}
    \includegraphics[height=7cm]{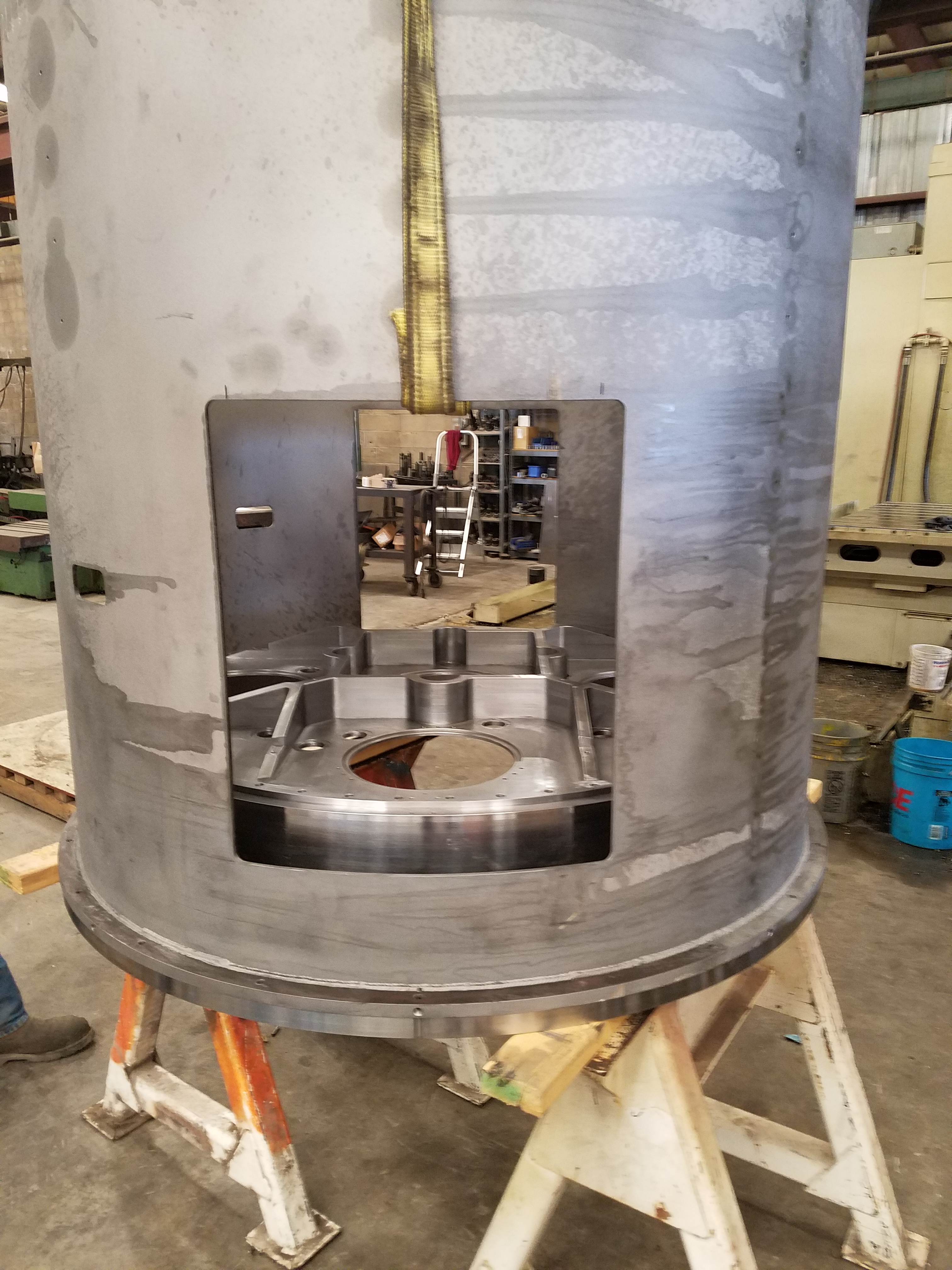}
   \end{tabular}
   \end{center}
   \caption 
   { \label{fig:CIoncorrector} Left: The Commissioning Instrument Focal Plate Assembly and Focal Plate System Mass (yellow), installed on the Mayall Telescope positioned just behind the corrector barrel (pink). The dashed black curve shows the approximate location of the focal surface. Right: The steel pieces for the Commissioning Instrument Focal Plate Assembly and Focal Plate System Mass. Most surfaces will be painted black.
}
   \end{figure}

The third component is the Focal Plane System Mass (FPSM). The CI FPSM is a large steel cylinder that protects the CI FPA and adds sufficient mass  to match the mass and moment of the DESI Focal Plane System. The left-hand panel of Fig. \ref{fig:CIoncorrector} shows the CI FPA and FPSM installed on the Mayall Telescope and positioned just behind the corrector barrel. The CI FPSM is shown in yellow and its front half has been cut away in order to show the CI FPA. The blue squares illustrate the access hatches that will allow members of the observing team to readily connect directly to the CI at the observatory. The right-hand panel shows a photograph of the steel pieces, prior to being painted.

All of the connecting faces of the Adapter Disk, Inner Cylinder of the FPA, and the Focal Plate are precision-ground to an RMS roughness of 0.8 microns. The faces are held flat to 30 microns and the parallel faces are held to within 15 microns. After we assemble the parts, they will be pinned to be concentric to within an uncertainty of 20 microns. 

\begin{figure}
   \begin{center}
   \begin{tabular}{c}
   \includegraphics[height=7cm]{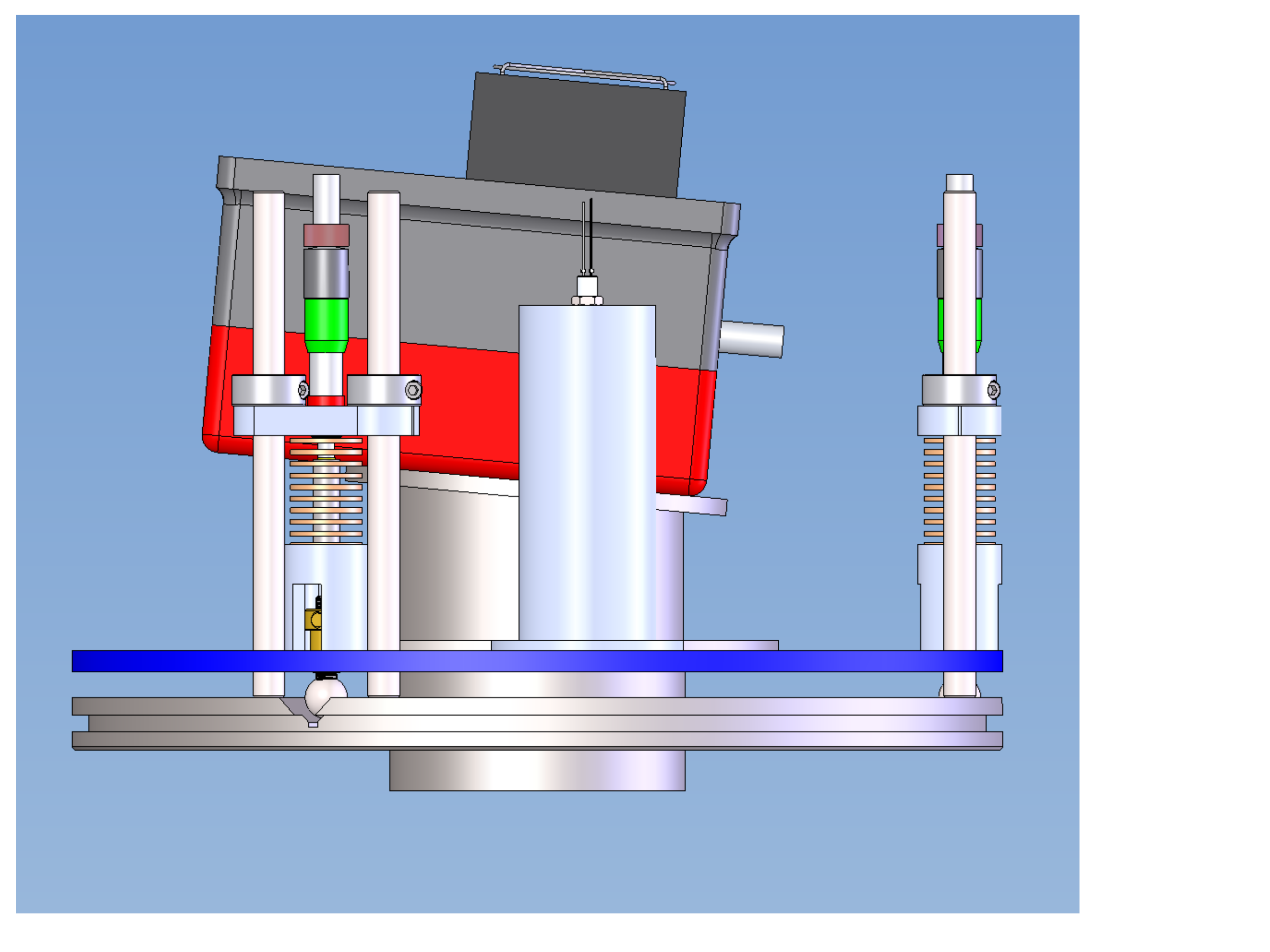}
   \includegraphics[height=7cm]{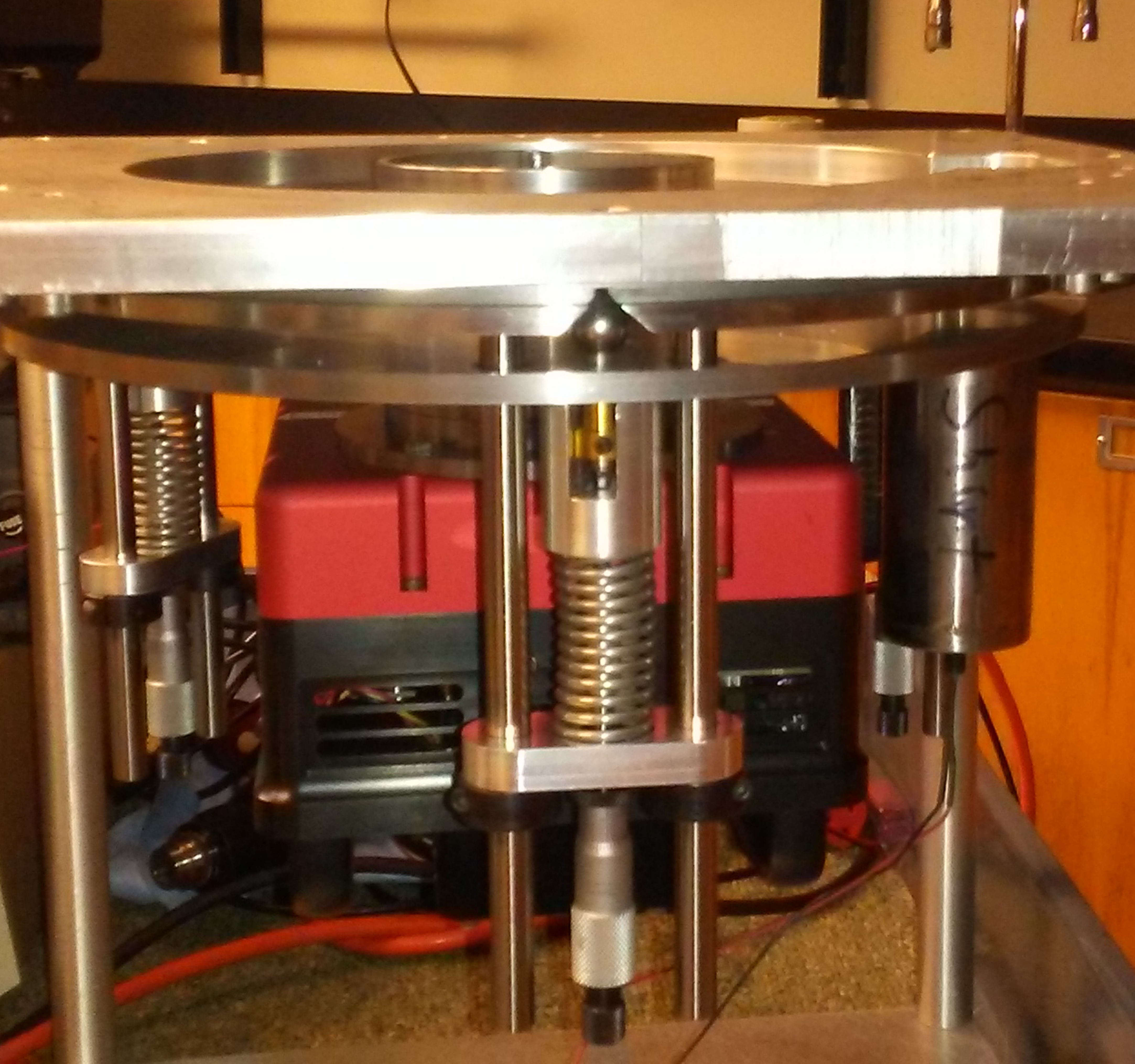}
   \end{tabular}
   \end{center}
   \caption 
   { \label{fig:TTF} The `Tip/Tilt/Focus' mount that will be used to mount each of the five SBIGs. The left-hand figure shows an illustration of the design, while the hand-hand figure is a picture of an SBIG camera attached to a prototype TTF mount, which is itself mounted to a test plate.
}
   \end{figure}

The SBIG cameras are mounted on the CI FPA using `Tip/Tilt/Focus' (TTF) mounts. The left-hand panel of Fig \ref{fig:TTF} shows an image of a camera in a TTF mount. The right-hand panel shows a photograph of a SBIG camera in a prototype TTF mount, which is itself mounted to a test plate. The camera is bolted to the steel connector in the center. This connection is flat for the center mount, and tilted for the outer mounts, in order to account for the curvature of the focal surface at those locations\footnote{Across the dimensions of the CCD, the height of the curved focal surface changes by $\pm$8 microns. Given our tolerance is 50 microns in terms of position relative the focal surface, we do not address this curvature beyond aligning the CCD to be tangent with the focal surface.}; this tilt is illustrated in the left-hand panel of Fig. \ref{fig:TTF}. The connector is attached to the TTF plate. The three degrees of freedom are controlled by three stainless steel balls that are in contact with the underside of the TTF plate. These three steel balls sit in v-grooves that are machined into the base plate. Spring towers hold the balls in place in the v-grooves. Movement of the plate is handled by a micrometer head that is connected to a spring-loaded, 80 threads per inch screw adjuster. For instance, adjusting all three by the same amount is equivalent to a focus correction.

The machined height of the fiducial mounts is adjusted based on the height of the focal surface where it will be mounted. The illuminated fiducials are threaded into place in their mounts. This interface allows $\pm$2mm of focus adjustment. This ensures that the four pinholes of the illuminated fiducial will be in focus. However, we do not tilt the fiducial mounts. Using laboratory tests, we have verified that the pinholes will be properly imaged by the FVC, despite their relative tilt. Each TTF mount will have one illuminated fiducial mount (the cylinder in Fig. \ref{fig:TTF}).

\subsection{Electrical Design}

 The DESI CI includes five cameras, five NUC computers, 22 illuminated fiducials, six temperature sensors, and an ethernet switch. These all require electrical power and ethernet connection for data transfer. Power is routed through a Raritan (model PX3-5145R\footnote{http://www.raritan.com/product-selector/pdu-detail/px3-5145r}) power distribution unit that allows individual control of power outlets via an ethernet connection.
 An ethernet switch will be plugged into the power strip and provide the data transfer. The Raritan unit and ethernet switch will connect to one power connection and one ethernet connection on the exterior of the FPSM. Each SBIG camera will be controlled by one Intel(R) NUC computer (model NUC5PPYH\footnote{https://ark.intel.com/products/87740/Intel-NUC-Kit-NUC5PPYH}), connected via USB. A DESI Petal Controller, produced by the University of Michigan, will control the 22 illuminated fiducials and the temperature sensors. The components that are not intended to align with the focal surface will be placed on an electronics shelf that will sit above the CI FPA and allow unstressed connections to all components and easy connection/disconnect. It is designed to minimize the likelihood of any disturbance of the precision aligned camera and fiducial components that plug into the components on the electronics shelf.

\subsection{Software Design}

\begin{figure}
   \begin{center}
   \begin{tabular}{c}
   \includegraphics[height=7cm]{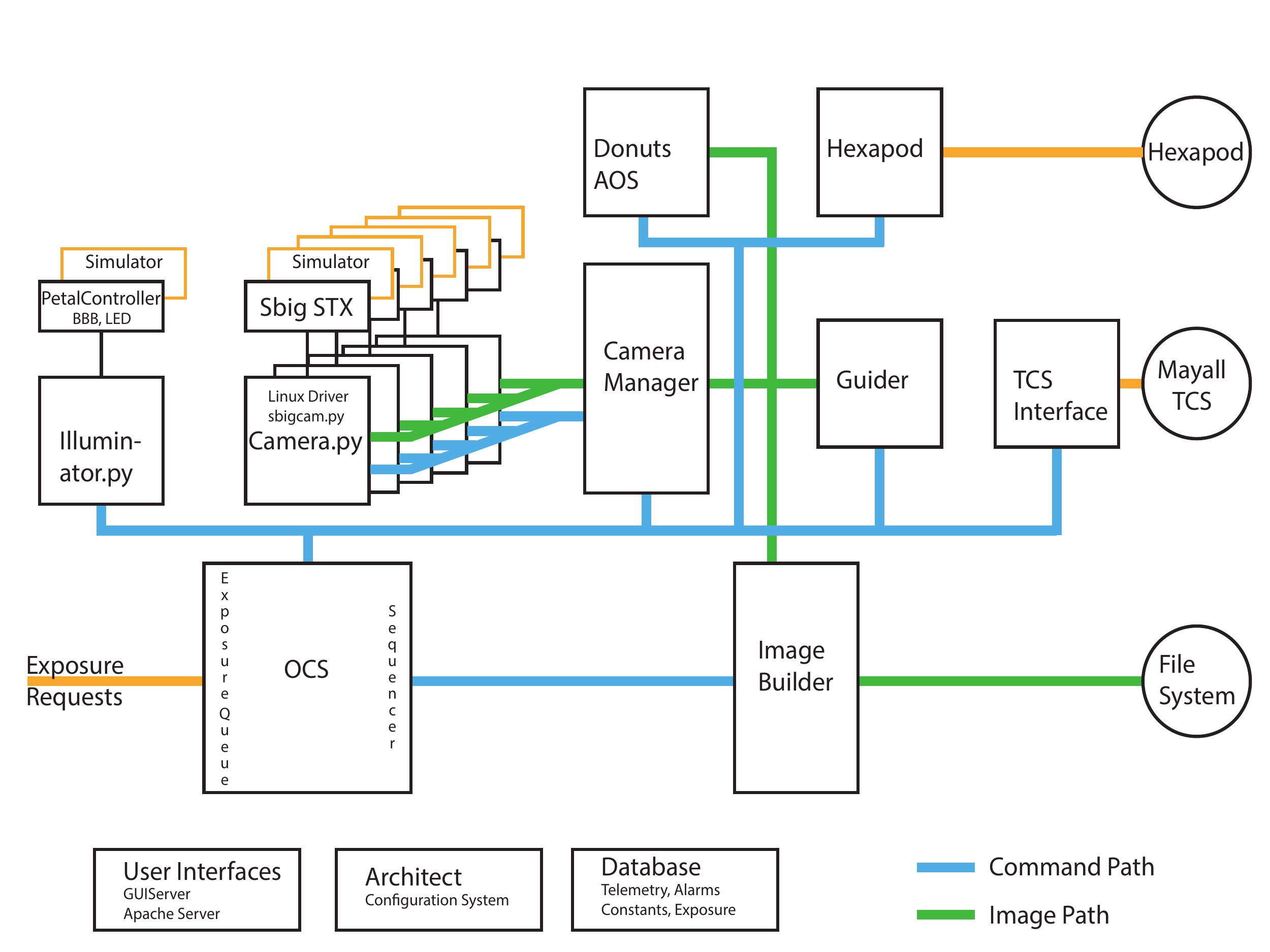}
   \end{tabular}
   \end{center}
   \caption 
   { \label{fig:software} The software design for the DESI Commissioning Instrument. The majority of the system is the same as the Instrument Control System that will be used for the full DESI system.
}
   \end{figure}

The DESI CI will be controlled with the DESI Instrument Control System (ICS). The design is shown in Fig. \ref{fig:software}. Laboratory tests to determine the readnoise and gain of the five SBIG cameras have been performed using the ICS camera control software. The diagram also shows that many vital components (e.g., the Hexapod control) will be controlled with the DESI ICS software for the CI. Operation of the CI will decrease risks related to the completion and performance of all of these systems. Further description of the DES ICS is included in a companion paper\cite{DESI_ICS}.

\section{OBSERVING CAMPAIGN}
\label{sec:obs}

The DESI Commissioning Instrument will be the first instrument used on the Mayall Telescope after it has been refurbished and upgraded\footnote{For instance, the whole top ring is being replaced.} with a new optical corrector. It will therefore perform crucial tasks that are required to demonstrate the basic functioning of the telescope. More specifically, its imaging capability will allow measurement of the image quality spanning the full DESI field of view. Further, given proper calibration, measurements of stellar photometry will characterize the $r$-band throughput of the DESI optical corrector. In what follows, we outline specific DESI commissioning tasks that will be completed using the CI.

\subsection{TCS and Guiding}

The Mayall Telescope Control System (TCS) is responsible for pointing the telescope at a given celestial coordinate. A new pointing model will be required after the installation of the DESI optical corrector. A first task is to develop this pointing model and verify its accuracy, e.g., by attempting to center the field of view on a given standard source. Secondly, we will test the stability of TCS tracking, e.g., by imaging stars over long exposures. Thirdly, we will use the CI to image dense star fields at high airmass and use the observations to verify that the Atmospheric Dispersion Corrector (ADC) works properly and the positions of objects are accurately predicted at these high airmass orientations. 

Once functionality of the TCS has been demonstrated, we will test the DESI guiding procedure. These tests will start by using one star on a single CCD, in order to verify basic functionality of the DESI guiding software, in coordination with the TCS. We will then test for differences in the quality of the results when using multiple stars per detector and when using all detectors. We will test the stability of the guiding over 20 minute exposures, which is the nominal exposure time for DESI science targets. The combination of this information will determine the nominal optimal procedure for guiding using the DESI GFA cameras.

\subsection{Hexapod}

The DESI hexapod\cite{DESI_INST} adjusts the position of the optical corrector in order to obtain optimal image quality. It allows corrections to the position of the optical corrector in terms of focus, tip, tilt, rotation, and de-center. The hexapod will be adjusted in order to obtain the optimal image quality. The nominal plan is for DESI to use an Active Optics System (AOS) that is similar to that used by the Dark Energy Survey\cite{Roodman14} (DES). Such a system characterizes the wavefront based on images of out of focus stars, and then moves the hexapod to achieve optimal image quality. Such images are taken between science exposures. The CI will commission the operation of the hexapod. A first test will be to simply achieve focus using one camera when pointed at zenith. Then, we will solve for the tip/tilt for the best focus across all five cameras.

With these initial measurements, we will characterize the image quality over a large portion of the field of view. We can compare this characterization to model predictions based on the characteristics of the as-built optical system (e.g., using the {\sc Zemax} software). To do so, we will intentionally mis-align the hexapod along all five movements and compare the measured images to those expected. We will do so over the full allowed range of motion in order to verify sufficient range in hexapod movements. This analysis will provide a basic commissioning of the DESI hexapod and its control software. Further, we will validate and improve on models of the optical performance of the as-built DESI optical corrector.

The next step will be to iteratively build a `look up table' (LUT) for the hexapod movements as a function of telescope orientation and temperature. A nominal LUT will have been created during the construction process, based on precise metrology of the telescope. The LUT will be verified and improved based on observations of fields with at least 30 stars per CCD observed in conditions with 1$^{\prime\prime}$.2 seeing and the variations that are observed as a function of temperature. In this way, we will empirically determine how well we can predict the hexapod settings based on the previous telescope position and the LUT.

At a minimum, an accurate LUT will minimize the amount of time during the DESI survey not taken by science exposures. Repeated tests will determine if a static LUT is sufficient for optimal image quality and the degree to which the AOS can improve quality during the course of a typical DESI exposure (rather than simply between exposures). The CI observing campaign will thus commission the DESI image optimization system and optimize the amount of telescope time that will be taken by DESI science exposures.

\subsection{Platemaker}
The DESI CI will be used to commission the software, `PlateMaker', that will be used to map between focal plane $X,Y$ and sky positions. This is a crucial component of the DESI project, as the FVC will be used to determine that fibers are at their correct $X,Y$ positions in order to observe astronomical targets. In order to do this mapping properly, an accurate model is required for both the distortion pattern over the DESI field of view at prime focus (influenced by the combination of the primary mirror and the optical corrector) and the distortion pattern over the DESI field of view as seen by the FVC (influenced by a combination of a different part of the optical corrector and the FVC optics).

PlateMaker will predict the astrometric solution for the five SBIG CCDs. We will test these predictions by observing fields with large numbers of stars with known accurate astrometry. The expected versus observed positions of stars on the SBIG CCD pixels will jointly test the model for the optics of the DESI corrector and the CCD geometry. The distortion pattern will vary with ADC angles. Therefore, we will observe over the full range of airmass for the DESI survey and thus confirm that PlateMaker accurately accounts for the ADC. The distribution of five SBIG cameras will sample both the central and radial portions of the DESI optical corrector and thus provide a good test of the distortion model.

Given an accurate mapping between ra,dec and SBIG pixel positions, we expect to consistently predict the FVC pixel position given the $X,Y$ value of an illuminated fiducial. This prediction requires knowledge of the limited portion of the optical corrector seen by the FVC and its own lens optics.
 Given the model for each, PlateMaker will predict the FVC pixel location for the 22 illuminated fiducials. Their positions will sample the DESI focal surface, with $X,Y$ positions illustrated in Fig. \ref{fig:CIFPA} (see also figure 3 of a companion proceedings\cite{ColesMetrology}). Their positions have been chosen so that a large enough portion of the focal surface is sampled in order to accurately determine the 16 coefficients required for the optical distortion model. Their relative positions will be known to within 10 microns\cite{ColesMetrology}. The DESI requirement is that PlateMaker maps between FVC position and ra,dec to an accuracy of 0.05 arcseconds for the DESI Focal Plane System. Comparisons between the accuracy of the PlateMaker astrometric predictions and FVC predictions will determine whether we are meeting this requirement and help diagnose any improvements needed to achieve the requirements, ahead of the installation of the full DESI Focal Plane System. Thus, the initial tests that will be done with the CI will help the project fulfill this ultimate requirement.

\section{CONCLUSIONS}
\label{sec:conclude}
The DESI Commissioning Instrument (CI) will commission the new Mayall optical system, thereby advancing the schedule of and reducing risk to the DESI program. We have described how the design of the DESI CI allows five commercial cameras and 22 illuminated fiducials to be precisely aligned on the prime-focus surface, while also matching the mass and moment of the DESI Focal Plane System. The DESI CI will perform vital commissioning tasks that range from verifying basic telescope function after its major-rebuild to verifying that DESI can accurately determine the focal plane position required to observe a target at a given sky position. The early testing of DESI components was demonstrated to be quite valuable during the ProtoDESI campaign\cite{ProtoDESI} and we expect similar value to come from the DESI CI campaign.


\acknowledgments     
 
This research is supported by the Director, Office of Science, Office of High Energy Physics of the U.S. Department of Energy under Contract No. 
DE-AC02-05CH1123, and by the National Energy Research Scientific Computing Center, a DOE Office of Science User Facility under the same 
contract; additional support for DESI is provided by the U.S. National Science Foundation, Division of Astronomical Sciences under Contract No. 
AST-0950945 to the National Optical Astronomy Observatory; the Science and Technologies Facilities Council of the United Kingdom; the Gordon 
and Betty Moore Foundation; the Heising-Simons Foundation; the National Council of Science and Technology of Mexico, and by the DESI 
Member Institutions: Aix-Marseille University;  Argonne National Laboratory; Barcelona Regional Participation Group; Brookhaven National Laboratory; 
Boston University; Carnegie Mellon University; CEA-IRFU, Saclay; China Participation Group; Cornell University; Durham University;  École Polytechnique 
Fédérale de Lausanne; Eidgenössische Technische Hochschule, Zürich;  Fermi National Accelerator Laboratory;  Granada-Madrid-Tenerife Regional 
Participation Group; Harvard University; Korea Astronomy and Space Science Institute; Korea Institute for Advanced Study; Institute of Cosmological  
Sciences, University of Barcelona; Lawrence Berkeley National Laboratory; Laboratoire de Physique Nucléaire et de Hautes Energies; Mexico Regional 
Participation Group; National Optical Astronomy Observatory; Ohio University; Siena College; SLAC National Accelerator Laboratory;  Southern Methodist University; 
Swinburne University; The Ohio State University; Universidad de los Andes; University of Arizona; University of California, Berkeley; University of California, 
Irvine; University of California, Santa Cruz; University College London; University of Michigan at Ann Arbor; University of Pennsylvania; University of Pittsburgh; 
University of Portsmouth; University of Rochester; University of Queensland; University of Toronto; University of Utah; University of Zurich; UK Regional Participation Group; Yale University. The authors are honored to be permitted to conduct astronomical research on Iolkam Du'ag (Kitt Peak), a mountain with particular significance to the Tohono O'odham Nation.  For more information, visit desi.lbl.gov.


\bibliography{report}   
\bibliographystyle{spiebib}   

\end{document}